%% file: main.tex
\documentclass[sigconf, 10pt]{acmart}
\usepackage{graphicx} 

\title{\LARGE See Behind Walls in Real-time Using Aerial Drones and Augmented Reality}

\author{Sikai Yang, Kang Yang, Yuning Chen, Fan Zhao, Wan Du}
\affiliation{
  \institution{Department of Computer Science and Engineering}
  \institution{University of California, Merced, USA}
  \city{}
  \country{}
}
\email{{syang126, kyang73, ychen372, fanzhao, wdu3}@ucmerced.edu}

\usepackage{eucal}
\usepackage{bm}
\usepackage{makecell}
\usepackage{algorithm}
\usepackage{lineno}
\usepackage{algpseudocode}
\usepackage{amsmath}
\usepackage{subfigure}
\usepackage{booktabs}
\usepackage{multirow}
\usepackage{enumitem}
\usepackage{flowchart}
\usepackage{float}
\usepackage{times}
\usepackage{url}
\usepackage{amsfonts}

\usepackage{amssymb}
\usepackage{xspace}
\usepackage{graphicx}
\usepackage{tabu}
\usepackage{diagbox}
\usepackage{gensymb}
\usepackage{caption}
\usepackage{subcaption}
\usepackage{xcolor}
\newcommand{\system}{$ARD^2$}

\renewcommand\footnotetextcopyrightpermission[1]{} 

\begin{document}
    \maketitle
    \pagestyle{plain}
    \input{abstract}

    \input{introduction}

    \input{background}
    \input{setup}

    \input{design}
    \input{FAQ}
    \input{experiment}
    \input{related}

    \input{conclusion}

\input{bb}
\end{document}

%% file: abstract.tex
\section*{Abstract}

This work presents \system,  a framework that enables real-time through-wall surveillance using two aerial drones and an augmented reality (AR) device. 
\system consists of two main steps: target direction estimation and contour reconstruction. 
In the first stage, \system leverages geometric relationships between the drones, the user, and the target to project the target’s direction onto the user’s AR display. 
In the second stage, images from the drones are synthesized to reconstruct the target’s contour, allowing the user to visualize the target behind walls. 
Experimental results demonstrate the system’s accuracy in both direction estimation and contour reconstruction.

%% file: introduction.tex
\section{introduction}
The ability to see through walls has long been a captivating idea. 
Such capability could provide immense value in applications like security and law enforcement \cite{1612798, borek2005through, anderson2013lightweight}. 
Existing through-wall surveillance methods primarily rely on wireless signals. 
For instance, X-AR~\cite{boroushaki2023augmenting} combines visual information with RF signals to render hidden RFID tags. 
However, these approaches face challenges when attempting to penetrate multiple layers of obstacles or operate over long distances \cite{DinaKatabi, wall_mimo, 4263932}.

\begin{figure}[ht]
    \begin{minipage}[t]{0.64\linewidth}
        \centering
        \includegraphics[width=\textwidth]{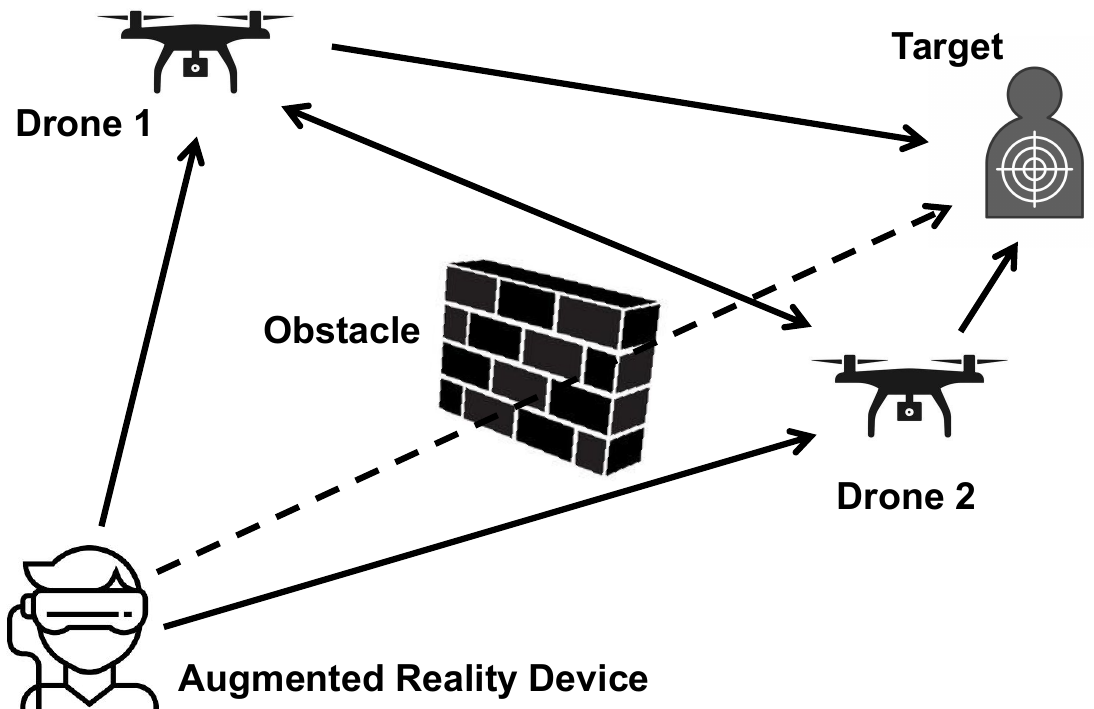}
         \vspace{-6mm}
        \caption{Application scenario}
        \label{fig_main}
    \end{minipage}
    \hfill
    \begin{minipage}[t]{0.34\linewidth}
        \centering
        \includegraphics[width=\textwidth]{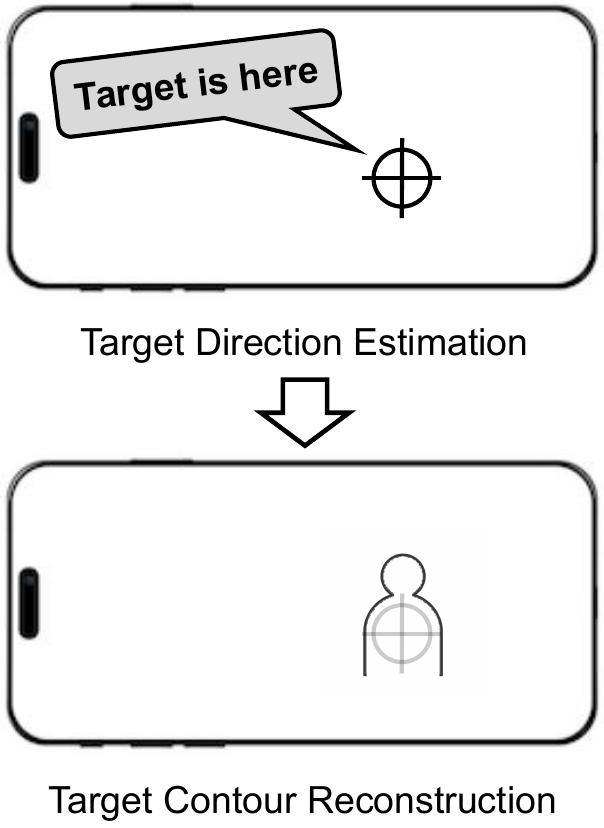}
         \vspace{-6mm}
        \caption{Two-step workflow}
        \label{fig_workflow}
    \end{minipage}
    \vspace{-4mm}
    \Description[bruh]{bruh bruh}
\end{figure}



In this paper, we introduce \system, a vision- and geometry-based through-wall surveillance solution (\textbf{A}ugmented \textbf{R}eality + Aerial \textbf{D}rones $\times$ \textbf{2}). 
Compared to wireless signal-based solutions, \system operates over longer distances and bypasses obstacles via relayed visual geometry.
As depicted in Figure \ref{fig_main}, \system leverages two aerial drones and an AR device to visualize targets located behind walls. 
The user operates by launching the two drones into the air while wearing an AR headset, with the target behind the wall. 
\system's goal is to render the target's direction and contour within the user's AR display. 
The process is divided into two main stages: target direction estimation and target contour reconstruction.

In the \textbf{direction estimation} stage, we determine the target's position within the AR by leveraging the geometric relationships between the user, drones, and the target.
The key insight is that the pixel-coordinate of an object in a photo correspond to its direction relative to the camera. 
By having the user and the two drones identify each other's directions, and the drones locating the target within their viewports, we can reconstruct the geometric relationship between the target and the user, and project it into the AR.
The result of this stage is also a pixel-coordinate—a dot in the user's AR indicating the target's direction.

Precise camera calibration can be challenging \cite{hartley2003multiple}.
To address this, we propose an on-site joint camera calibration scheme that reduces the overhead of calibrating both the AR device and drone cameras. 
The core idea is based on the fact that the three angles between the user and the two drones must always sum to 180°.
We use the deviation from this sum as a loss function to jointly fine-tune the camera parameters. 
By ensuring the AR device and drones capture sufficient images of each other, we can iteratively minimize camera distortion, enabling accurate target direction tracking.

The \textbf{contour reconstruction} stage builds on the previous one.
In this stage, we use the images captured by the two drones to synthesize the target image that the user would have seen if the walls were not present.
Specifically, we focus on reconstructing only the contour of the target, rather than a full-color image.
Although several solutions exist for synthesizing images from multiple views, we face the significant challenge of working with only two images—far fewer than required by most existing few-shot view synthesis methods \cite{mildenhall2020nerf, yang2023freenerf, wu2024reconfusion}.

Fortunately, we leverage application-specific conditions to help bridge this gap.
Existing view synthesis methods are designed to handle a wide range of target types, but we can focus on specific categories of interest.
For instance, real-time tracking of live targets, such as humans or animals, is far more critical than tracking inanimate objects like chairs.
Since \system might potentially serve professionals with specific objectives, we can narrow the range of potential target types.
This allows us to use a digital 3D model of the desired target type to generate synthetic visual data from various angles.
With this generated data, we can train a specialized neural network for \system to accurately reconstruct the target's contour from the two drone-captured images.

We conduct real experiments to evaluate \system.
The results show a 2.7° error in target direction estimation and a 94\% accuracy in target contour reconstruction.

In conclusion, the contributions of this work are as follows:
\begin{itemize}[leftmargin=*]
    \item We design an on-site joint camera calibration scheme that requires minimal effort for camera calibration.
    \item We design an on-site joint camera calibration scheme to calibrate cameras with low effort.
    \item We carefully account for application conditions and design a data augmentation framework to support few-shot target contour reconstruction.
    \item We conduct real experiments to evaluate the system and assess its potential for real-world applications.
\end{itemize}



    
    
    


    


%% file: background.tex
\section{Background} \label{sec_background}

\textbf{Camera Geometry: }
Camera geometry refers to the mathematical framework used to model how cameras capture and project 3D scenes onto a 2D image plane. 
This involves understanding the relationship between a camera's physical components, such as the lens, sensor, and focal length.
Closely related to camera geometry is camera calibration, which involves estimating both intrinsic and extrinsic parameters to accurately model the camera's behavior. 
Specifically, intrinsic parameters include focal length, the principal point (where the optical axis intersects the image plane), and lens distortion coefficients that account for optical distortions like barrel and pincushion effects. 
Accurate calibration is crucial for 3D reconstruction, augmented reality, and multi-view geometry, where precise knowledge of these parameters ensures proper interpretation and projection of visual data.
With precise camera geometry, it is practical to infer the direction of an object using its pixel-coordinate within a photo.

\textbf{View Synthesis: }
View synthesis is a technique in computer graphics and computer vision that involves generating new images or perspectives of a scene from existing views. 
By understanding the spatial relationships between objects and the camera's viewpoint, view synthesis can interpolate or extrapolate additional angles.
This process often leverages depth information, multi-view images, or 3D models to produce realistic visuals. 
View synthesis is critical for applications like virtual reality, augmented reality, and 3D rendering, where immersive experiences require seamless transitions between viewpoints. 
Advancements in neural networks and light field technology are further pushing the boundaries of view synthesis accuracy and realism.

%% file: setup.tex
\section{Scenario Setup}
We first introduce the application scenario of \system, including several important technical configurations.

\subsection{Potential Application and Advantage}
Seeing behind walls in real time can provide significant assistance to various applications, such as security and law enforcement \cite{1612798, borek2005through}.
Another example is hunting \cite{anderson2013lightweight}, where hunters would be able to see their prey behind tree lines, which is not possible with the naked eye.
Such real-time visual feedback can offer a substantial tactical advantage.
With the system rendering the direction and contour of the target in AR glasses, hunters can aim and shoot through the woods in the estimated direction of the prey.
This tactical advantage not only improves hunting efficiency but also enhances the safety of hunters due to the unidirectional surveillance capability of the system.
In this scenario, long distances and dense layers of foliage may attenuate wireless signals, rendering traditional methods impractical.

\subsection{Technical Configurations}
To enhance understanding, we clarify several technical configurations, including system input and output formats, as well as prerequisites.

\subsubsection{Hardware}
\system requires two drones with cameras on board and an AR device with an integrated camera.
Fortunately, both aerial drones and AR devices typically come with integrated cameras nowadays \cite{9289536, AR_survey}.

\subsubsection{System Prerequisites}
\system requires the AR camera to detect both drones within its field of view, while both drones must also detect the AR and the other drone, as well as the target.
Such line-of-sight condition is usually met in outdoor scenarios, as the drones are airborne.

Additionally, \system does not assume that targets are stationary.
For instance, in a hunting scenario, prey may be running.
In this case, we treat each time spot as a complete data sample, with the AR camera and two cameras on the drones capturing a video frame simultaneously.

\subsubsection{Walls}
\system can work on walls, as well as any obstacles that block the line-of-sight between the user and the target, as long as the line-of-sight conditions are satisfied for user-to-drones, drone-to-drone, and drones-to-target.

\subsubsection{System Raw Inputs}
At each time step, \system only takes three photos as input, from the AR camera and two drone cameras.
No depth information or distance measurement is required by \system.

\subsubsection{Target Identification and Contour Extraction} \label{sec_preprocess}
Photos can contain various and numerous objects, either relevant or irrelevant to the task.
Before the main work flow of \system, it requires the AR and drones to be identified with in the photos.
This identification step can be accomplished using many existing solutions \cite{yolo, ren2016faster, deeplab, dxz_cv}.
Furthermore, the target needs to be both identified by the system, and confirmed by the user as the tracking subject.
Once the target is confirmed, the system will continuously track the same target based on visual similarity and continuity until the target disappears from the drones' view.
The final pre-processing involves extracting the contour of the target from the images captured by the drones, which can also be achieved using various existing methods \cite{canny1986computational, xie2015holistically}.
As a result, the final inputs to the main workflow include:

\begin{itemize}[left=0pt]
    \item The pixel-coordinates of two drones within the AR camera.
    \item The pixel-coordinates of the AR, the other drone, and the target within both drone's camera.
    \item The contour of the target extracted from both drone's view.
\end{itemize}

\subsubsection{Two-step Outputs}
As shown in Figure \ref{fig_workflow}, \system first performs target direction estimation to determine the target's direction relative to the user.
This direction can be visualized as a dot in the user's AR.
Next, \system reconstructs the contour of the target, and overlays it upon the dot in the user's AR.

%% file: design.tex
\section{Design}
The workflow of \system consists of two primary steps: target direction estimation and target contour reconstruction.
Specifically, target contour reconstruction is built upon the outcomes of target direction estimation.

\subsection{Target Direction Estimation}
The target direction estimation module aims to find the direction of the target within the user's AR reference.
It visually notifies the user by rendering a dot within the AR interface.

\subsubsection{Drone-AR-Drone Triangle}
We first establish a triangle between the AR and two drones.
As introduced in Section \ref{sec_background}, the direction of a detected object can be inferred from its pixel-coordinate within a photo.
Using the directions of the AR and two drones detected within each other's scope, we can create a well-defined triangle and represent it using vectors in the AR's reference system.

\subsubsection{AR-Drones-Target Tetrahedron} \label{tetrahedron}
Images captured by the two drones also imply the target direction, which can be represented as a linear combination of the vector from the AR to the drones, the vector between the two drones, and the cross product of these two vectors.
Consequently, within the AR reference, we can extend the direction vectors from the drones to the target, until they intersect or reach the minimal distance.
This intersection can be considered as the target location.
Thus, we successfully complete the last missing edge of the AR-drones-target tetrahedron, enabling us to render the target direction as a dot in the AR.

\subsubsection{On-site Joint Camera Calibration}
The foundation of the target direction estimation module is camera geometry, which may not always be readily available.
Many online camera specifications are incomplete, and third-party technical reviewers often lack access to comprehensive internal specification data, resulting in imprecise information.

In the case that camera specifications are not fully available and to ensure that \system can be deployed with random combinations of off-the-shelf drones and AR cameras, we propose an on-site joint camera calibration scheme.
The core insight is that the three angles formed between the user and the two drones must always sum up to 180°.
We use the difference between this sum and 180° as the loss function for optimization to fine tune the camera parameters jointly.
As long as the user's AR  and the drones capture sufficient images of each other, we can progressively estimate accurate camera specifications to support reliable target direction tracking.

\subsection{Target Contour Reconstruction}
Based on the direction estimation results, we further utilize the images captured by the drones to reconstruct the contour that the user would have seen if there were no walls.

\subsubsection{Challenge}
\begin{figure}[ht]
    \hfill
    \begin{minipage}[t]{0.6\linewidth}
        \centering
        \includegraphics[width=\textwidth]{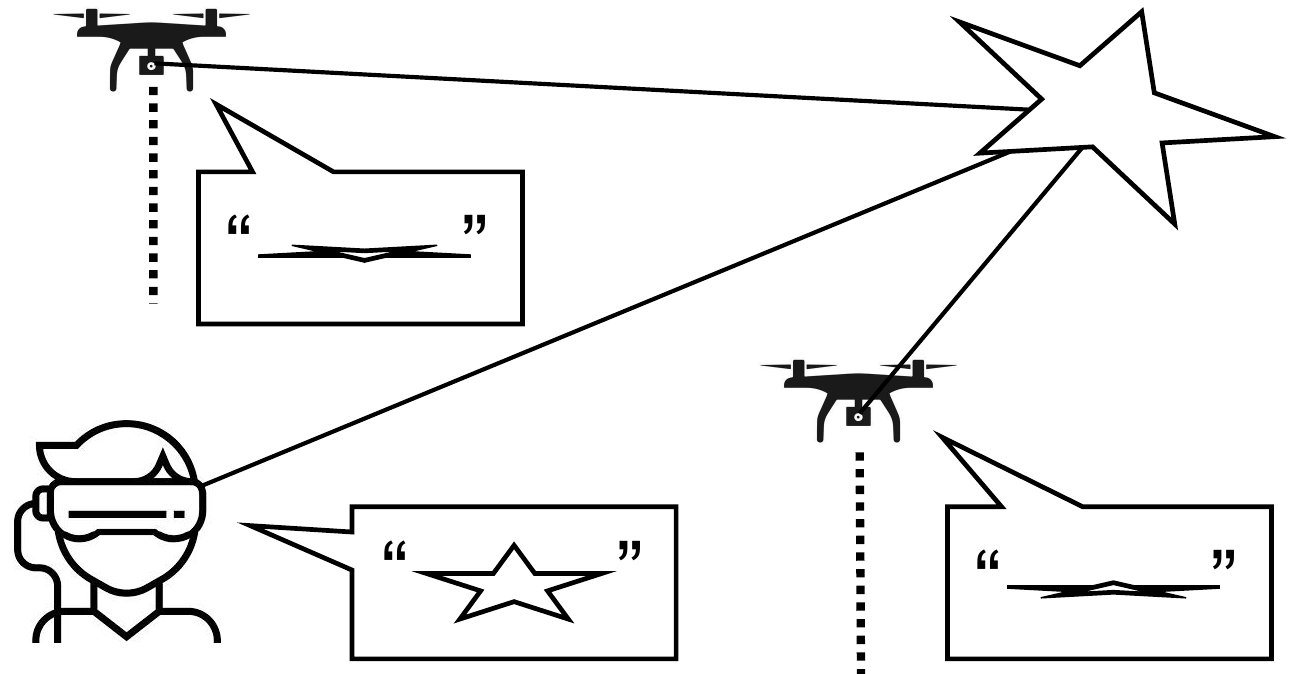}
         \vspace{-6mm}
        \caption{Extreme case illustration}
        \label{fig_extreme}
    \end{minipage}
    \hfill
    \begin{minipage}[t]{0.36\linewidth}
        \centering
        \includegraphics[width=\textwidth]{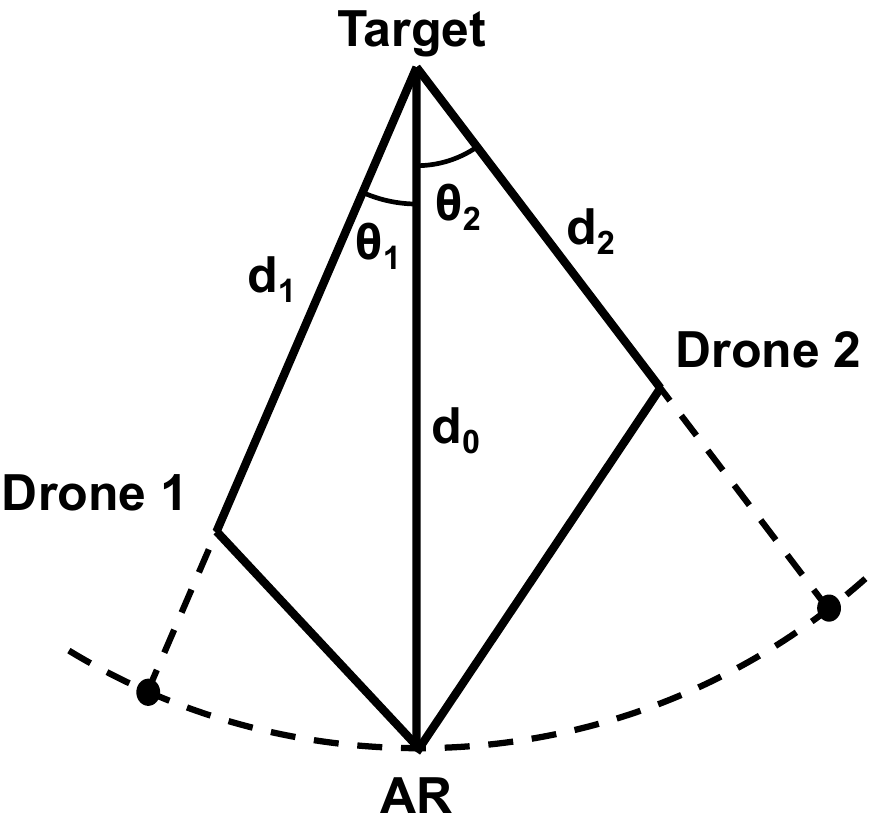}
         \vspace{-6mm}
        \caption{Input contour scaling}
        \label{fig_scale}
    \end{minipage}
    \hfill
    \vspace{-5mm}
    \Description[bruh]{bruh bruh}
\end{figure}

Few-shot view synthesis has been well studied \cite{kerbl20233d, yang2023freenerf, wang2023sparsenerf}. 
or instance, NeRF variants and 3DGS can reduce the amount of required training images to 24.
However, in our scenario, we are limited to only two drones providing two camera angles, which is insufficient for existing view synthesis solutions \cite{mildenhall2020nerf, yang2023freenerf, wu2024reconfusion}.
Moreover, our application can be theoretically impossible in certain extreme cases if we rely solely on camera information.
For example, as shown in Figure \ref{fig_extreme}, a flat object can be carefully positioned so that the drones only capture its edge.
This results in a lack of necessary information to infer the correct result.
This issue primarily arises from the unbridgeable altitude gap between the user and the drones.
Therefore, it is crucial to apply external knowledge to assist the system, such as assumptions about drone placement and potential target type information.

\subsubsection{Contour Reconstruction Neural Network}
As introduced in Section \ref{sec_preprocess}, the input of the contour reconstruction stage is the contours detected in the drones' view.
We first use the distances from the target to the AR and drones to scale the input contour images to simulate the same photo distances.
As illustrated in Figure \ref{fig_scale}, the scaling factors are $\frac{d_0}{d_1}$ for drone 1, and $\frac{d_0}{d_2}$ for drone 2.
The ratio of $\frac{d_0}{d_1}$, $\frac{d_0}{d_2}$ can be calculated from the AR-drones-target tetrahedron from Section \ref{tetrahedron}.
Since the drones may capture the photos at a tilted attitude, we also rotate the input contours to align them parallel to the AR's horizon (X-axis), using the drones' orientations.
Finally, we scale the input contour images to $60\times60$ resolution.


Since the drones are launched by the user and perform through-wall surveillance tasks, we assume that they are closer to the user than the target, and thus have a small camera angle difference to the AR (e.g., 30°).
Based on this assumption, we aggregate the input contours into one, with a weight of the cosine of their camera angle difference to the AR ($cos \theta_1$ and $cos \theta_2$ in Figure \ref{fig_scale}).
We then employ five convolutional layers to output the contour within the AR.
This aggregation framework also provides flexibility for the system, allowing for the inclusion of more than two drones if needed.

\subsubsection{Simulation-based Data Augmentation}
\system is likely to be utilized by professionals with specific surveillance tasks focused on certain target types, which allows us to narrow the scope.
To leverage this, we utilize online-available 3D target models (e.g., human models with adjustable skeletons) to digitally simulate the application scenario.
We can also simulate various camera positions to generate sufficient visual data from different angles.
Using the generated data (3600 data samples from 3D human model), we pre-train a general model.
Finally, to bridge the gap between simulated data and reality, we fine-tune the model using real data.

%% file: FAQ.tex
\subsection{Potential Technical Concerns}

\subsubsection{Could One Drone be Sufficient?}
The Synthetic Aperture Radar (SAR) \cite{sar} concept moves a single radar platform along a flight path to synthesize a large aperture for high-resolution imaging.
A similar solution to our application involves flying a drone around for multiple camera angles.
However, such concept typically assumes stationary targets; otherwise, the target location would change while the drone is scanning. 
To support moving target tracking, we regard each time spot as a complete data sample, in which the drones and the target can all be considered static.



\subsubsection{Distance Measurement?}
Distance measurement is not required by \system, since we only utilize angles and orientations to reconstruct the user-target direction.
For the contour reconstruction, we rely on the relative ratio between the drone-target distance and the user-target distance to scale the contours captured by the drones.


\subsubsection{Why Real-time Augmented Reality?}
Real-time augmented reality provides high-volume visual information that could enable great tactical advantage.
Based on the detection results of aerial drones, traditional radio-based conveyance transforms the visual information from the drones to linguistic or digital form, which largely decreases information density and can have minute or second-level delay.
In contrast, AR interface directly receives full visual information and display it in the user's reference in real-time, which allows the user to react to situations with millisecond-level latency.

\subsubsection{Why Contour Only?}
\system only reconstructs the contour of the target, rather than the full image with content and colors.
Overlaying an imaginary figure in the AR may divert users' attention and potentially degrade user experience \cite{buchner2022impact, ibili2019effect}.
Conversely, a target contour would be enough for the user to identify the target and observe its activities.

%% file: experiment.tex
\section{Preliminary Results}
We conducted real experiments to evaluate the performance of \system, including target direction estimation accuracy, contour reconstruction accuracy, and their combined end-to-end performance.
To collect data efficiently, we used phone cameras instead of drone and AR cameras.
We used building balconies to simulate aerial drone altitudes.
Real drone cameras can be affected by flight vibration, leading to motion blur and focus loss.
However, it can be mitigated by existing vibration control methods \cite{ali2017stability, verma2020active, beltran2022active}.

\begin{figure} [ht]
    \subfigure[Drone 1]{\includegraphics[width=0.215\linewidth]{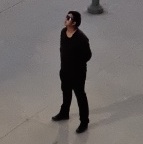}}
    \hfill
    \subfigure[Drone 2]{\includegraphics[width=0.225\linewidth]{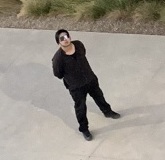}}
    \hfill
    \subfigure[User's view]{\includegraphics[width=0.22\linewidth]{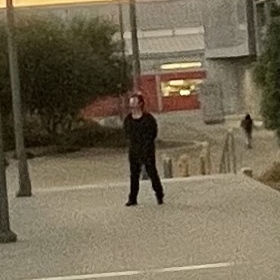}}
    \hfill
    \subfigure[Final]{\includegraphics[width=0.22\linewidth]{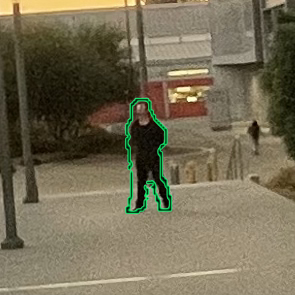}}

    \vspace{-3mm}

    \subfigure[Input 1]{\includegraphics[width=0.22\linewidth]{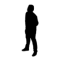}}
    \hfill
    \subfigure[Input 2]{\includegraphics[width=0.22\linewidth]{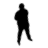}}
    \hfill
    \subfigure[G. Truth]{\includegraphics[width=0.22\linewidth]{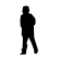}}
    \hfill
    \subfigure[Inference]{\includegraphics[width=0.22\linewidth]{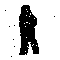}}

    \vspace{-5mm}
    \caption{Raw image, input, output, and final results}
    \vspace{-6mm}
    \label{fig_final}
    \Description[bruh]{bruh bruh}
\end{figure}

\subsection{Implementation}
We collected 75 data samples, with different user-drones-target geometric settings.
We mainly used an iPhone 12, an iPhone 13 pro, and an iPhone 15 pro max to represent the AR camera and two drones' camera, respectively.
In our experiment, the drone-user-target angle ranges from approximately 10° to 45°.
User-target distance ranges from 10m to 100m.
Drone altitude ranges from 0 to 20m.
To collect ground truth data, we did not have a wall between the user and the target.
Target-related information within the user's AR camera was used for evaluation purposes only.
We included two types of target: human figure and a bobcat statue in the campus, as shown in Figure \ref{fig_final} and \ref{fig_bobcat}.
Due to the lack of the 3D model of the bobcat statue, we only enable data augmentation for the human figure target.
For contour reconstruction neural network, we allocate 60 for training and the rest for testing.

\subsection{Direction Estimation Accuracy}
\begin{figure}[ht]
    \begin{minipage}[t]{0.5\linewidth}
        \centering
        \includegraphics[width=\textwidth]{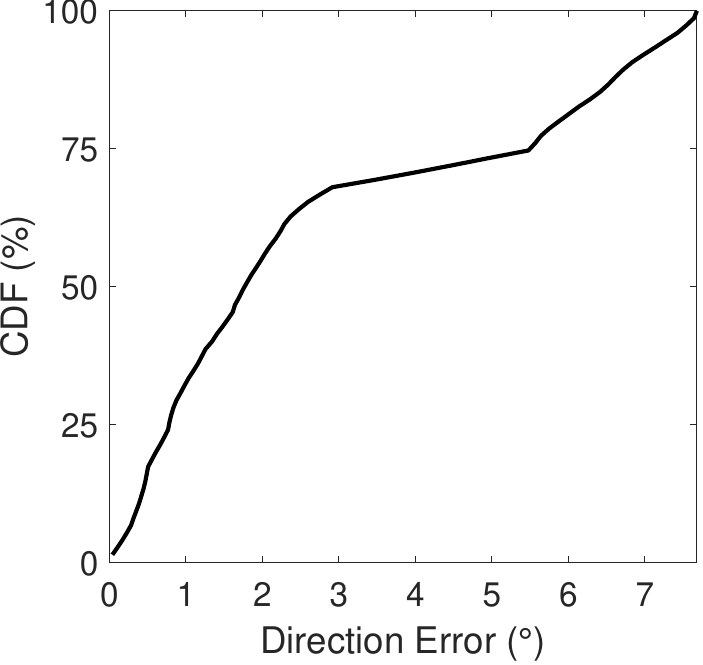}
         \vspace{-7mm}
        \caption{Target direction estimation accuracy}
        \label{fig_expm_direction}
    \end{minipage}
    \hfill
    \begin{minipage}[t]{0.46\linewidth}
        \centering
        \includegraphics[width=\textwidth]{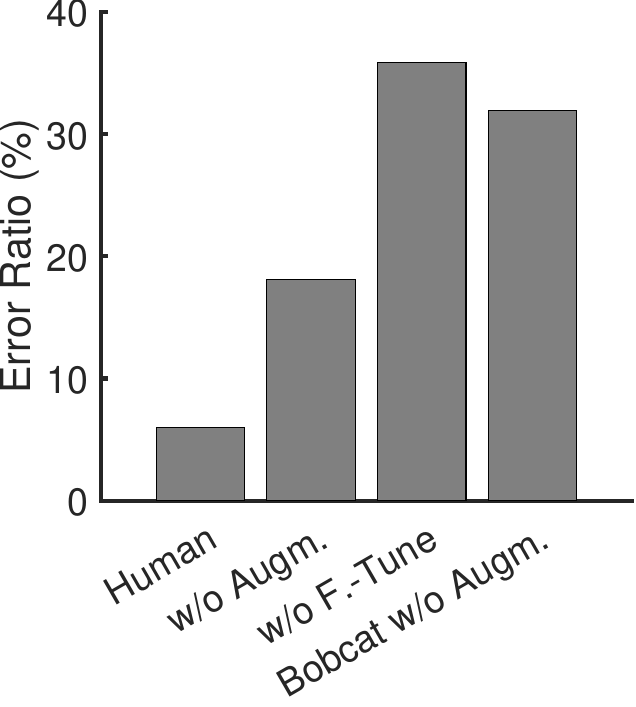}
         \vspace{-7mm}
        \caption{Target contour reconstruction accuracy}
        \label{fig_expm_contour}
    \end{minipage}
    \vspace{-5mm}
    \Description[bruh]{bruh bruh}
\end{figure}

The average target direction estimation error is 2.69°.
This accuracy is primarily attributed to the robust camera geometry.
Figure \ref{fig_expm_direction} shows the distribution of the error.
There is a noticeable concentration of relatively high error around 7°, which is likely due to under-synchronization between cameras when taking photos together.

\subsection{Contour Reconstruction Accuracy}
\begin{figure}[ht]
    \begin{minipage}[t]{0.3\linewidth}
        \centering
        \includegraphics[width=\textwidth]{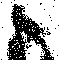}
         \vspace{-6mm}
        \caption{Inferred contour}
        \label{fig_13}
    \end{minipage}
    \hfill
    \begin{minipage}[t]{0.3\linewidth}
        \centering
        \includegraphics[width=\textwidth]{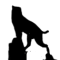}
         \vspace{-6mm}
        \caption{Ground truth}
        \label{fig_13gt}
    \end{minipage}
    \hfill
    \begin{minipage}[t]{0.3\linewidth}
        \centering
        \includegraphics[width=\textwidth]{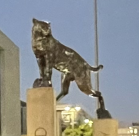}
         \vspace{-6mm}
        \caption{Bobcat statue}
        \label{fig_bobcat}
    \end{minipage}
    \vspace{-3mm}
    \Description[bruh]{bruh bruh}
\end{figure}

We evaluate the contour reconstruction accuracy by calculating the error ratio, which is defined as the amount of unmatched pixels between the output and ground truth contour image, divided by the total amount of pixels within the ground truth contour, i.e., the contoured area.
Visualized contour results are shown in Figure \ref{fig_final}(h) and \ref{fig_13}.

Figure \ref{fig_expm_contour} shows the statistics of contour reconstruction accuracy for different models and targets.
For the human figure target, when \system enables both data augmentation and fine-tuning, it achieves 5.96\% error ratio.
When data augmentation is disabled, it achieves 18.12\% error ratio.
When fine-tuning is disabled, error ratio rises to 35.84\%.
These results implies that both data augmentation and fine-tuning are essential.
Without data augmentation, the model struggles to learn well with limited data availability.
Meanwhile, without fine-tuning, the model fails to effectively bridge the gap between simulation and reality, likely due to factors like body shape and clothing.
Lastly, the bobcat statue has a error ratio of 31.94\%, which is higher than human figure without data augmentation.
This increased error is likely attributable to the greater visual complexity of the bobcat statue's contour.

\subsection{End-to-end Result}
Based on the inferred contour, we perform basic noise reduction, resolution up-scaling, and smoothing.
Finally, using the results from both direction estimation and contour reconstruction, we visualize the contour within the AR's view, as demonstrated in Figure \ref{fig_final}(d).
It is important to note that \system can still render this green contour even if the target is block by a wall.
The green contour slightly deviates from the actual figure of the target, which be attributed to the direction estimation error.

\subsection{System Latency}
We implemented \system's algorithm on an Alienware M18 R1 laptop equipped with an RTX 4090 GPU and 13th Gen Intel i9-13980HX 2.20 GHz CPU.
On average,\system spends 0.119ms on the direction estimation stage, and 1.048ms on the contour reconstruction stage.
Although mobile platforms have fewer computation resources, these results indicate that \system is likely to be able to achieve real-time performance on such platforms.

%% file: related.tex
\section{Related Work}


\textbf{Few-shot view synthesis.}
Traditional methods like NeRF~\cite{mildenhall2020nerf} require more than one hundred input images to obtain the output.
To tackle this issue, recent studies attempt to achieve few-shot synthesis.
FreeNeRF~\cite{yang2023freenerf} introduces a dynamic frequency control module for few-shot NeRF optimization.
ReconFusion~\cite{wu2024reconfusion} uses diffusion models to synthesize additional views and jointly trains Zip-NeRF for sparse-view rendering.
3DGS~\cite{kerbl20233d} applies Gaussian splatting to increase the efficiency of reconstruction.
However, none of them can tackle the sparse input size of two.
Multiple UAVs can contribute diverse visual hints, which are precious for reconstructing the objects or scenes\cite{schmuck2017multi}. 
However, it needs multi-location samples to reconstruct the point cloud. 
Using two UAVs, \system fully considers practical real-world assumptions and proposed a 3D simulation-based data augmentation framework to reconstruct the target image.

\textbf{Augmented reality.}
Augmented reality (AR) is a technology that enhances the real-world environment by overlaying digital content such as images, sounds, or information. 
Unlike virtual reality, which creates an entirely immersive experience, AR blends the physical world with interactive, computer-generated elements in real-time. 
There are two categories of AR implementation: Video See-Through (VST) and Optical See-Through (OST).
VST captures the real world through cameras and displays the augmented view on screens, which is a mature technical path and has been adopted by most hardware, e.g., Meta Quest, Apple Vision Pro. OST uses transparent displays that allow users to see the real world directly while overlaying virtual content, e.g., Meta Orion glasses~\cite{meta2024orion}.
SLAM-share~\cite{dhakal2022slam} conducts accurate 3D map merging across multiple AR devices, improving the shared AR experience for users.
In this work, \system discusses a new application that enables outdoor see-through-obstacle capability for AR users.



%% file: conclusion.tex
\section{Conclusion}

In this paper, we present \system, a system that enables real-time through-wall surveillance using two aerial drones and an AR device. The system effectively estimates the target direction and reconstructs the target’s contour in the user’s AR view.
The evaluation shows a 2.69° direction estimation error and 5.96\% contour reconstruction error, demonstrating accurate performance with minimal latency.